\documentclass[10pt,a4paper]{article}

\usepackage{amsmath}
\usepackage{amsfonts}
\usepackage[parfill]{parskip}
\usepackage{subfig}
\usepackage{graphicx}
\usepackage{vmargin}
\usepackage{color}
\usepackage{changebar}
\usepackage[round]{natbib}
\usepackage[noblocks]{authblk}
\definecolor{darkblue}{rgb}{0,0,0.6}
\usepackage[colorlinks=true,citecolor=darkblue]{hyperref}

\newcommand{\be}{\begin{equation}}
\newcommand{\ee}{\end{equation}}
\newcommand{\ba}{\begin{equation} \begin{aligned}}
\newcommand{\ea}{\end{aligned} \end{equation}}
\newcommand{\ddt}[1]{\frac{\mathrm{d}#1}{\mathrm{d}t}}
\newcommand{\dint}[1]{\mathrm{d}#1}
\newcommand{\myvec}[1]{ \mathbf{#1} }
\newcommand{\mymat}[1]{ \mathbf{#1} }
\newcommand{\mygmat}[1]{ \boldsymbol{#1} }
\newcommand{\mygvec}[1]{ \boldsymbol{#1} }

\title{\bf Algebraic moment closure for population dynamics on discrete structures}
\author{Thomas House}
\date{}

\begin{document}

\maketitle

\begin{abstract}
\noindent{}Moment closure on general discrete structures often requires one of
the following: (i) an absence of short closed loops (zero clustering); (ii)
existence of a spatial scale; (iii) \textit{ad hoc} assumptions. 
Algebraic methods are presented to avoid the use of such assumptions for
populations based on clumps, and are applied to both SIR and macroparasite
disease dynamics.  One approach involves a series of approximations that can be
derived systematically, and another is exact and based on Lie algebraic
methods. 
\end{abstract}

\section{Introduction}

\subsection{Motivation}

In modelling biological systems, we typically have (at least) three objectives:
enhanced scientific understanding of the system in question; the ability to
answer counterfactual questions relating to possible perturbations of the
system; and a framework for working with data to estimate values of important
parameters statistically.  Having a low-dimensional model is of benefit for
each of these aims since such simple models are typically analytically more
transparent and numerically more efficient than complex high-dimensional
models.

Increasingly, models of infectious disease transmission take place on a network
to represent the heterogeneities that exist in contacts between
hosts~\citep{Danon:2011}.  This is a major benefit for model realism, however
in the situation where $N$ distinct individuals are in one of $\Sigma$
epidemiological states then $\Sigma^N$ variables are needed to describe the
system state. A low-dimensional description of such dynamics is therefore
essential if we seek understanding, optimisation of intervention strategies or
statistical inference in the large-$N$ regime.

\subsection{Background}

Many different approaches exist to enable dimensional reduction of
high-dimensional models. Perhaps the most commonly used is Kirkwood's
superposition principle \citep{Kirkwood:1942}. In the context of
biology, this approach has been applied to infectious diseases
\citep{Keeling:1999}, including rapid modelling of emerging disease threats
\citep{Ferguson:2001}. Such equations are intended to be used in the case of
relatively homogeneous populations, but with significant transitivity in the
pairwise interactions. Here, transitivity is used to imply the
presence of a significant number of triangles in the network of interactions
between individuals, which would be called `clustered' in network theoretical
language that we adopt here. While these moment closures are often in
excellent agreement with simulation, they lack \textit{a priori} justification
meaning we cannot be clear about exactly when they will fail.

One class of populations where there is potentially significant
transitivity in the network of interactions and where some analytical progress
can be made are spatial models. Here individuals are located at points in space
and the interaction strength between them decays as the distance between them
increases, with $L$ being the natural length scale of the interaction. 
Here, the pioneering work of \citet{Ovaskainen:2006}, and \citet{Cornell:2008}
demonstrated how asymptotically exact results could be obtained to avoid these
concerns by perturbative expansion in an inverse spatial length scale $L^{-1}$.

For non-clustered but heterogeneous populations, work by \citet{Ball:2008} has
also showed how asymptotically exact epidemic equations can be derived that are
at worst of dimensionality $2N$, with more sophisticated limit theorems
\citep{Decreusefond:2012} showing that the four-dimensional equations of
\citet{Volz:2008} can be derived -- and indeed the dimensionality of these
equations can be further reduced to one \citep{Miller:2011,Miller:2012}. Such
insights can be applied to non-spatial clustered populations if there is a
local-global distinction, to provide sets of ODEs whose dimensionality, while
often large, can be independent of $N$ \citep{House:2010,Volz:2011,Ma:2013}.

At the same time, there are innovative approaches to non-exact clustered moment
closure, including: projection operator methods \citep{Dodd:2007}; analytic
consideration of asymptotic early growth \citep{House:2010a}; maximum entropy
methods~\citep{Rogers:2011}; non-independent Bernoulli
trials~\citep{Taylor:2012}; and \textit{a priori} distributions
\citep{Kiss:2012}. These often significantly improve closure performance, but
remain essentially \textit{ad hoc}.

\subsection{Algebraic methods}

This paper considers an algebraic approach to moment closure for clustered
populations based on `clump' models, but generalisable to other local-global
structures. This can take the form either of a series of approximations or an
exact solution based on Lie algebras. As such it provides a potential new route
to rigorous moment closure for clustered discrete structures. We will first
introduce some general methods for time-inhomogeneous ODE systems, then will
apply these to SIR epidemics and macroparasite ecology.

\section{Algebraic solution of time-inhomogeneous ODEs}

\subsection{General theory}

Suppose we have a time-inhomogeneous set of ODEs that are written in matrix /
vector form as
\be
\ddt{\myvec{p}} = \mymat{M}(t) \myvec{p}
\text{ .} \label{kf}
\ee
We will seek solutions to this equation of the form
\be
\myvec{p}(t) = {\rm e}^{\mymat{Z}(t)} \myvec{p}(0) \text{ .}
\label{expz}
\ee
The case where $\mymat{M}$ does not depend on time was considered in the
context of epidemic dynamics by \citet{Keeling:2008}, and offers significant
computational advantages. There are a variety of approaches to the solution of
the more general case outlined in a particularly comprehensive paper by
\citet{Wilcox:1967}, the conventions from which will in general be used here.
We will consider two such approaches.

\subsection{The Magnus expansion}

The first approach we will consider is a series expansion introduced by
\citet{Magnus:1954}, which has previously been used to solve equations
like~\eqref{kf} applied to population dynamics~\citep{Ross:2012}. Here
we start by re-writing~\eqref{kf} as
\be
\ddt{\myvec{p}} = \mu \mymat{M}(t) \myvec{p}
\text{ .} \label{mkf}
\ee
Substituting~\eqref{expz} into~\eqref{mkf} then gives the relation
\be
\mu \mymat{M}(t) = 
\int_0^1 \mymat{G}(t,u) \dint u \text{ ,} \qquad \text{for} \quad
\mymat{G}(t,u) = {\rm e}^{u \mymat{Z}(t)} \dot{\mymat{Z}}(t)
{\rm e}^{- u \mymat{Z}(t)} \text{ ,} \label{zint}
\ee
where we use a dot to denote derivatives with respect to time.  Since this
integral cannot be performed directly in the most general case, we consider a
series
\be
\mymat{Z}(t) =
{\sum_{k=1}^{\infty} \mu^k \mygmat{\Omega}_k(t)}
\text{ .} \label{ms}
\ee
We will refer to the truncation of the infinite sum at $\mygmat{\Omega}_m$ as
the Order $m$ approximation.  Substituting~\eqref{ms} into~\eqref{zint} and
equating terms at the same order in $\mu$ (after which point we set $\mu=1$
without loss of generality) gives a sequence of matrices, with the first three
being
\ba
\mygmat{\Omega}_1(t) & = \int_{t_1=0}^{t} \mymat{M}(t_1) \dint{t_1} \text{ ,}\\
\mygmat{\Omega}_2(t) & = \frac{1}{2}
\int_{t_1=0}^{t} \int_{t_2=0}^{t_1}
[\mymat{M}(t_1), \mymat{M}(t_2)] \dint{t_1} \dint{t_2} \text{ ,}\\
\mygmat{\Omega}_3(t) & = \frac{1}{6}
\int_{t_1=0}^{t} \int_{t_2=0}^{t_1} \int_{t_3=0}^{t_2} 
\left(
[[\mymat{M}(t_1),\mymat{M}(t_2)], \mymat{M}(t_3)]+
[[\mymat{M}(t_3),\mymat{M}(t_2)], \mymat{M}(t_1)]
\right)
\dint{t_1} \dint{t_2} \dint{t_3}
\text{ .} \label{me}
\ea
Here the matrix commutator $[\mymat{A},\mymat{B}]:= \mymat{A}\mymat{B} -
\mymat{B}\mymat{A}$.  The convergence of such an expansion is
considered by \citet{Blanes:1998}, who argue that convergence will occur for
times $t$ such that
\be
\| \mygmat{\Omega}_1(t) \| < \xi \text{ ,} \label{magconv}
\ee
and provide the numerical estimate $\xi \approx 1.086869$. Note that this is a
sufficient but not necessary condition for convergence so matrices with larger
norm can still be part of a convergent series.

\subsection{Lie algebra methods}

\label{sec:lie}

Now suppose that we can write
\be
\mymat{M}(t) = \sum_{i=1}^{l} a_i(t) \mymat{H}_i \text{ ,} 
\label{lie}
\ee
where the $\{\mymat{H}_i\}_{i=1}^{l}$ is a set of linearly independent matrices
obeying
\be
[\mymat{H}_i, \mymat{H}_j] = \sum_{k=1}^{l} {\zeta_{ij}}^k \mymat{H}_k \text{
,} \qquad {\zeta_{ij}}^k\in \mathbb{C} \text{ .}
\label{liealg}
\ee
Such matrices form a representation of a \textit{Lie algebra}, and we will now
leave the range $1, \ldots l$ for indices $i,j,\ldots$ implicit for notational
compactness.  We then follow \citet{Wilcox:1967} and note that $\mymat{G}(t,u)$
as in~\eqref{zint} obeys the following system of differential equations:
\be
\frac{\partial \mymat{G}}{\partial u} = [\mymat{Z}(t),\mymat{G}(t,u)] \text{ ;}
\qquad
\mymat{G}(t,0) = \dot{\mymat{Z}}(t) \text{ .} \label{liedes}
\ee
In general, these equations are not an improvement on~\eqref{zint}, however
when the Lie algebra decomposition~\eqref{lie} is possible, we can expand
other matrices in the same basis
\be
\mymat{Z}(t) = \sum_i \alpha_i(t) \mymat{H}_i \text{ ,} \qquad
\mymat{G}(t,u) = \sum_i \mathcal{A}_i(t,u) \mymat{H}_i \text{ ,}
\label{liezg}
\ee
and use the linear independence of the $\mymat{H}_i$ to solve for
$\mymat{Z}(t)$.

\section{Clump-based epidemics}

\subsection{Model definition}

A highly influential paper by \citet{Ball:1997} considered epidemics on
networks with `two levels of mixing' in which the population is split into $m$
fully connected `clumps' (or `cliques' in network theoretical terminology) of
size $n$, so that $N=m\times n$. The large-population limit is taken with
$m\rightarrow \infty$ at constant finite $n$; this leads to a natural
interpretation of the population structure as a network with the `small worlds'
property of significant clustering \emph{within} the clumps, but small mean
shortest path lengths due to the random nature of links \emph{between} clumps.

One of the most common applications of this work is to households, which are
an important locus of disease transmission and also a natural unit for the
collection of epidemiological data. Then $n$ is the size of a household, which
will typically be several orders of magnitude smaller than the whole population
size $N$. Variable household sizes can easily be included in this framework,
but here we keep $n$ fixed to simplify the analysis.

While the analysis of \citet{Ball:1997} mainly concerned the early and late
behaviour of an epidemic in a population of clumps / households, later work
modelled temporal dynamics and intervention strategies using `self-consistent'
equations \citep{Ghoshal:2004,Dodd:2007,House:2008, House:2009}.  For SIR
epidemics these take the form of $(n+1)(n+2)/2$ ODEs, which hold in the
large-$N$ limit,
\ba
\ddt{}{p}_{s,i}
& = \gamma \left( -i p_{s,i} + (i + 1) p_{s,i+1} \right) \\
& \quad + \tau \left( -s i p_{s,i} + (s + 1) (i - 1) 
  p_{s+1,i-1} \right) \\
& \quad + \beta I(t) \left( -s p_{s,i} + (s + 1)
  p_{s+1,i-1} \right) \text{ .}
\label{sc}
\ea
Here $p_{s,i}(t)$ is interpreted as the proportion of clumps with $s$
susceptible and $i$ infectious individuals at time $t$, meaning that
$p_{s,i}=0$ if $s+i>n$. $\gamma$ is the recovery rate, $\tau$ is the rate of
within-clump transmission and $\beta$ is the rate of between-clump
transmission.  It will be convenient to define quantities
\be
{\sigma}_{s,i} := s/n \text{ ,}\quad
{\iota}_{s,i} := i/n \text{ ,}\quad
{v}_{s,i} = si/n \text{ .} \label{vecdef}
\ee
Like $p_{s,i}$, these have two indices and so are 2-tensors, however since the
possible number of states is finite, we can easily concatenate columns to work
with vectors $\myvec{p}(t)$, $\mygvec{\sigma}$, $\mygvec{\iota}$ and
$\myvec{v}$.  While $p_{s,i}$ is a proportion rather than a probability, since
clumps are not created or destroyed in the model $\myvec{p}$ has very similar
properties to a probability vector and its elements sum to unity.  In vector
notation, we have the following expressions for the proportions of the
population susceptible and infectious:
\be
S(t) := \mygvec{\sigma}\cdot \myvec{p}(t) \text{ ,} \qquad
I(t) := \mygvec{\iota}\cdot \myvec{p}(t) \text{ .} \label{sidef}
\ee
The definitions~\eqref{vecdef} and~\eqref{sidef}, together with~\eqref{sc},
define the high-dimensional system for closure. Of course, such a system
already achieves independence of $N$, but accounting for local clustering can
still lead to high dimensionality.  Our task will be to derive a set of ODEs
whose dimensionality does not depend on $n$ or $N$.

\subsection{Standard moment closure}

We now consider existing moment closure approaches.  Both maximum entropy and
the use of Ans\"{a}tze would lead to the assumption that $p_{s,i}$ is a
multinomial probability mass function for $n$ trials with probabilities $S$ and
$I$ of observing susceptible and infective on each trial. Kirkwood's
superposition principle would simply involve the direct assumption that the
probability of a pair of distinct individuals in the same clump being
susceptible-infective is equal to the product $S \times I$, meaning that all
standard approaches would lead to the substitution
\be
\mathbf{v}\cdot \mathbf{p} \rightarrow (n-1) SI \text{ ,}
\ee
which can be combined with~\eqref{si} to make a closed system. We call this the
`mean-field' model.

\subsection{Order 1 algebraic moment closure}

\label{sec:order1}

We now consider how algebraic approaches such as the Magnus expansion can be
applied to the question of moment closure. We first note that two manipulations
of the full self-consistent field equations~\eqref{sc} are possible.  The first
of these is to write them in
the form of~\eqref{kf},
\be
\ddt{\myvec{p}} = (\mymat{H} + I(t) \mymat{K}) \myvec{p}
\text{ .} \label{hk}
\ee
We can then apply the Order 1 Magnus expansion given in~\eqref{me} to the 
system~\eqref{hk}, assuming that $I(t)$ is a known function, to give
\be
\mygmat{\Omega}_1(t) = \mymat{H}t + \mymat{K} 
\int_{t_1=0}^{t} I(t_1) \dint{t_1} =:
\mymat{H}t + \mymat{K}C(t) 
\text{ .} \label{om1}
\ee
The second manipulation of~\eqref{sc} involves taking the inner product of
these equations with each of $\mygvec{\sigma}$ and $\mygvec{\iota}$ to give an
unclosed pair of ODEs that generalise the standard SIR model,
\ba
\ddt{S} & = - \beta S I - \tau \mathbf{v}\cdot \mathbf{p} \text{ ,}\\
\ddt{I} & = \beta S I - \gamma I + \tau \mathbf{v}\cdot \mathbf{p} \text{ .}
\label{si}
\ea
We can then substitute~\eqref{om1} into~\eqref{ms} to obtain a matrix
exponential approximation to $\mathbf{v}\cdot \mathbf{p}$, provided the
cumulative incidence of infection $C(t)$ is known.  Fortunately, $C(t)$ can be
seen to obey a simple ODE from its definition.  This leads to a set of three
ODEs approximating the full dynamics,
\ba
\ddt{S} & = - \beta S I - \tau \mathbf{v}\cdot
\mathrm{exp}\left(\mymat{H}t + \mymat{K}C(t)\right)
\mathbf{p}(0) \text{ ,}\\
\ddt{I} & = \beta S I - \gamma I + \tau \mathbf{v}\cdot
\mathrm{exp}\left(\mymat{H}t + \mymat{K}C(t)\right)
\mathbf{p}(0) \text{ ,}\\
\ddt{C} & = I(t) \text{ .}
\label{si1}
\ea
It is worth spending some time on consideration of the intuitive
interpretation of this closure regime, which is equivalent to the assumption
that $\myvec{p}(t) = \myvec{q}(u)|_{u=t}$ , where
\be
{\frac{\mathrm{d}{\myvec{q}}}{\mathrm{d}u}} =
\left(\mymat{H} + \frac{C(t)}{t} \mymat{K}\right) \myvec{q}(u) \text{ ,}
\qquad \myvec{q}(0) = \myvec{p}(0) \text{ .} \label{hku}
\ee
Note that $\mymat{H}$ corresponds to the events that take place within the
clump -- recovery of infectives and transmission between members of the same
clump -- and $\mymat{K}$ corresponds to a susceptible within the clump becoming
infective due to infectives external to the clump. Therefore, we can interpret
the approximation as assuming that the system state at time $t$ is equivalent
to the one that would be obtained by running an epidemic with clumps each of
whose susceptibles spontaneously become infective at constant rate $\beta
C(t)/t$ integrated over the time period $u\in [0,t]$.

Therefore, the approximation is expected to be completely accurate if all
infection external to the clumps is time-homogeneous; increased variability in
$I(t)$ over the epidemic is therefore expected to lead to a less accurate
approximation.

\subsection{Higher order moment closure}

The main attraction of the Magnus series is that it can be iterated to produce
higher-order results. In particular, the Order 2 and 3 matrices are
\be
\mygmat{\Omega}_2(t) = 
\frac{1}{2} \left(D_1(t) - D_2(t)\right)\mymat{L}
 \text{ ,} \qquad
\mygmat{\Omega}_3(t) = 
\frac{1}{6} F_1(t)\mymat{Q}_1 - F_2(t)\mymat{Q}_2
 \text{ .}
\ee
Here we have used matrices
\be
\mymat{L} := [\mymat{H}, \mymat{K}] \text{ ,} \quad
\mymat{Q}_1 := [\mymat{L}, \mymat{H}] \text{ ,}\quad
\mymat{Q}_2 := [\mymat{L}, \mymat{K}] \text{ ,}
\ee
and functions satisfying ODEs
\begin{gather}
\ddt{D_1} = C(t) \text{ ,}\quad
\ddt{D_2} = t I(t) \text{ ,}\quad
\ddt{E} = I(t) C(t) \text{ ,}\nonumber \\
\ddt{F_1} = 2D_2(t) - \frac{1}{2}t^2I(t)-D_1(t) \text{ ,}\quad
\ddt{F_2} = E(t) + \left(D_2(t)-D_1(t)\right) I(t) \text{ .}
\end{gather}
The benefits of all approximations considered above in terms of dimensionality
reduction are displayed in Table~\ref{dimtab}.

\begin{table}[h]
\begin{center}
\begin{tabular}{|c|c|}
\hline
\textbf{Model} & \textbf{Dimensionality}\\
\hline
Full & $(n+1)(n+2)/2$ \\
Order $1$ & $3$ \\ 
Order $2$ & $5$ \\ 
Order $3$ & $8$ \\ 
Mean Field & $2$ \\ 
\hline
\end{tabular}
\end{center}
\caption{Dimensional benefits of algebraic closure for the epidemic clump model.\label{dimtab}}
\end{table}

\subsection{Numerical results}

\begin{figure}[h]
\centering
\scalebox{0.75}{\resizebox{\textwidth}{!}{ \includegraphics{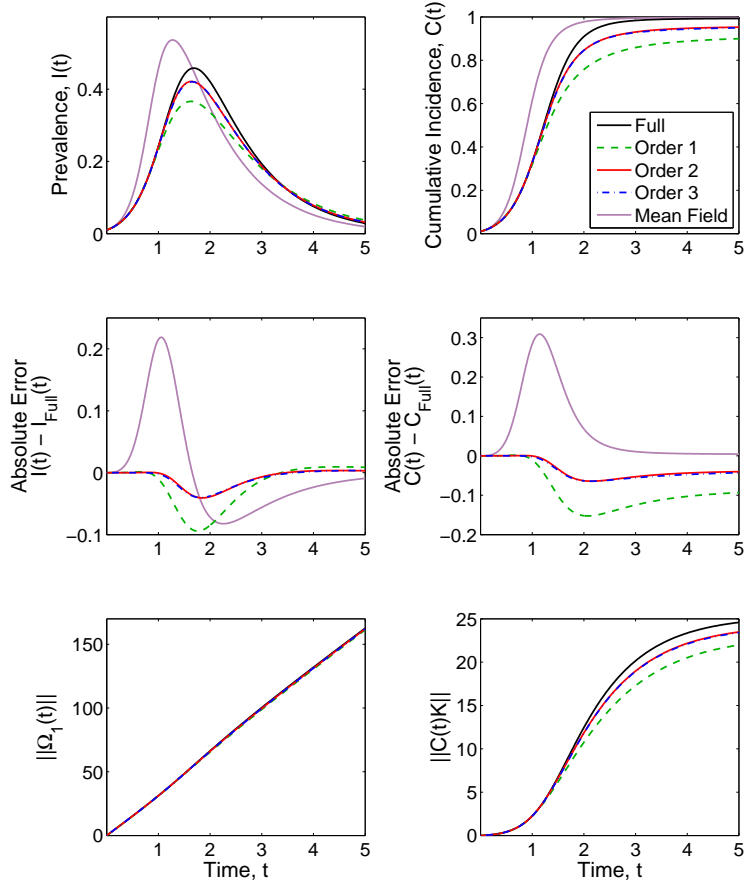} }}
\caption{Epidemic dynamics for $n=10$, $\beta = 3/2$, $\gamma = 1$, $\tau =
1/2$ and 1\% of the population initially infectious.
\label{magfig}}
\end{figure}

\begin{figure}[h]
\centering
\scalebox{0.75}{\resizebox{\textwidth}{!}{ \includegraphics{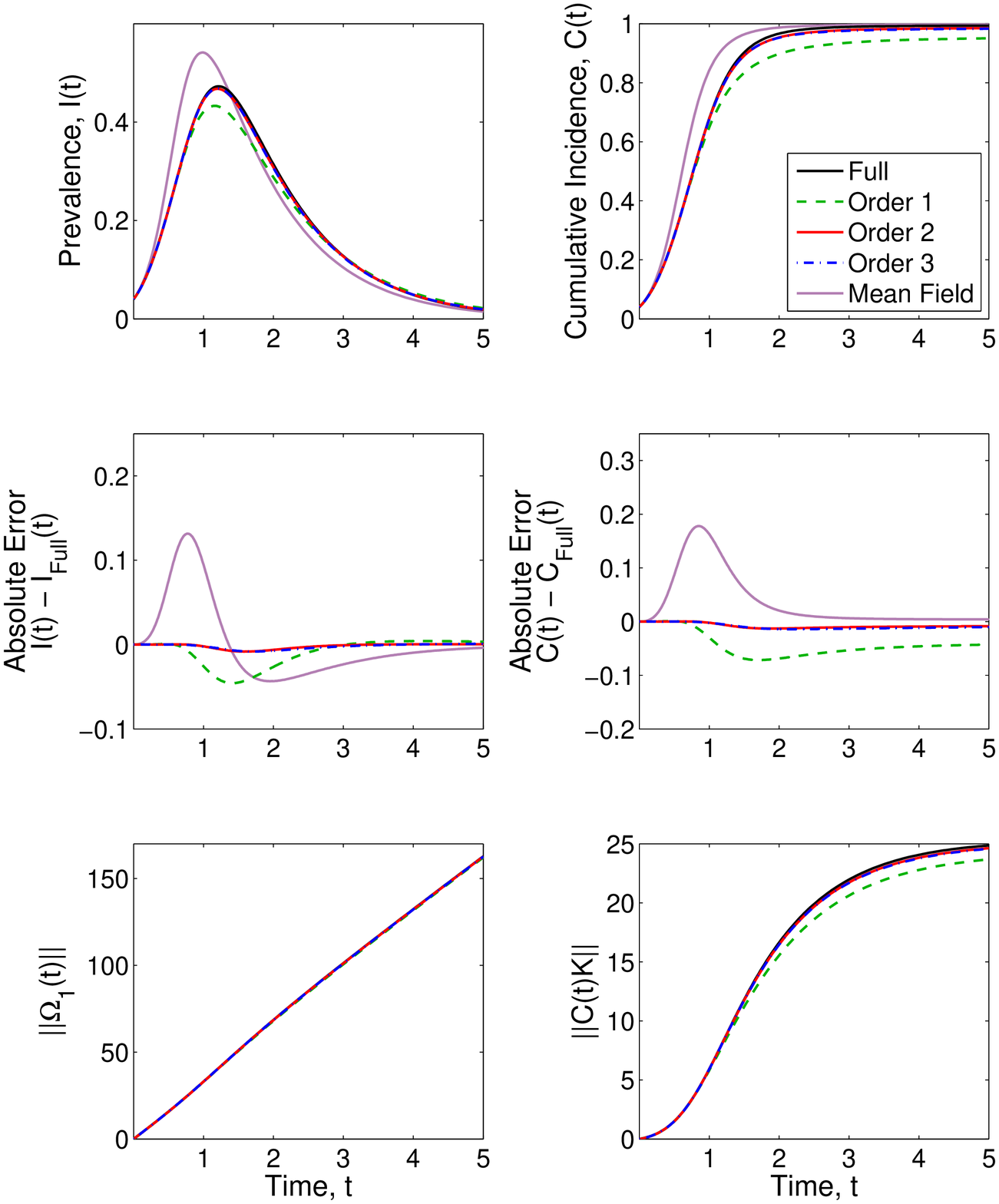} }}
\caption{Epidemic dynamics for $n=10$, $\beta = 3/2$, $\gamma = 1$, $\tau =
1/2$ and 4\% of the population initially infectious.
\label{magfig2}}
\end{figure}

The results of integrating the closed equations at different orders, making use
of the EXPOKIT routines \citep{Sidje:1998}, is given in Figures~\ref{magfig}
and~\ref{magfig2}.  This na\"{i}ve method is used to test the
accuracy of the approach and in practice it would be better to use one of
several advanced numerical integration schemes designed to work efficiently
with the Magnus series~\citep{Iserles:1999,Blanes:2002,Blanes:2013}.

For short time periods, the norm $\|\Omega_1(t)\|$ is less than the value given
in Eq.~\ref{magconv} that is known to be sufficient for convergence of the
Magnus series onto the exact solution \citep{Blanes:1998}. This value is
quickly exceeded, however, and numerical results suggest that while the series
does converge, it does so onto something that is not an exact solution
to~\eqref{hk}, raising the question of how accurate an approximation is
obtained. Since there is seldom much difference between the Order 2 and Order 3
systems, we will tend to consider the Order 2 approximation from now on.

Comparing Figures~\ref{magfig} and~\ref{magfig2} we see that the agreement
between approximation and exact results is very good for initial infectious
prevalence of 4\%, but not as good for 1\% -- although in both scenarios the
approach clearly outperforms the mean-field model for early times.  We can
interpret these results in light of the discussion of \S{}\ref{sec:order1}
above -- if $I(t)$ is too variable over time, we do not expect the
approximation to be good since then the solution of~\eqref{hku} is expected to
be a further from the solution of~\eqref{hk}.  The smaller initial condition
involves more variability in $I(t)$ over time than the larger, with analogous
(but more complex) argumentation holding at higher orders.

\begin{figure}[h]
\centering
\scalebox{0.75}{\resizebox{\textwidth}{!}{ \includegraphics{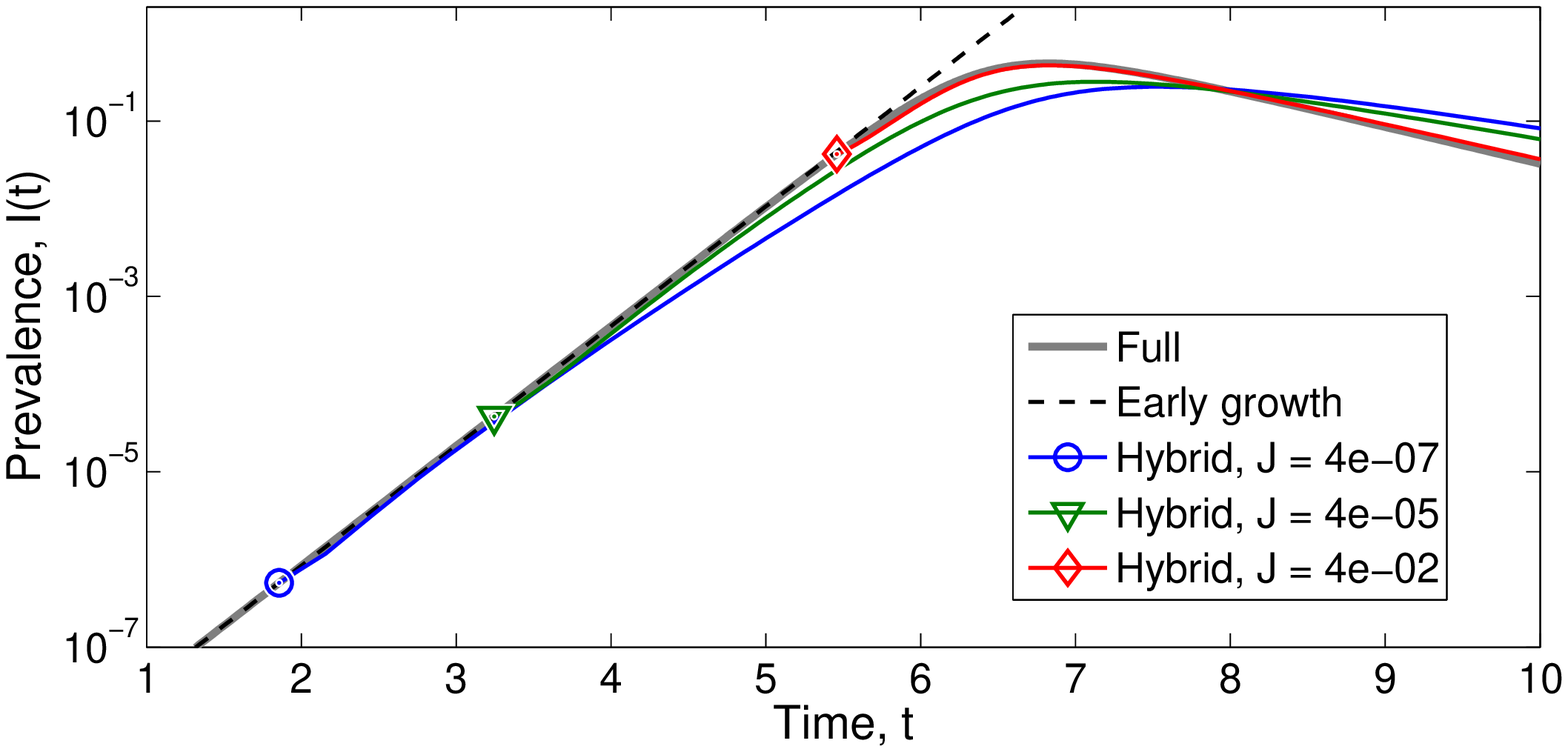} }}
\caption{Performance of hybrid approaches for different crossover prevalences
$J$, with $n=10$, $\beta = 3/2$, $\gamma = 1$, $\tau = 1/2$. \label{hyb}}
\end{figure}

\begin{figure}[h]
\centering
\scalebox{0.75}{\resizebox{\textwidth}{!}{ \includegraphics{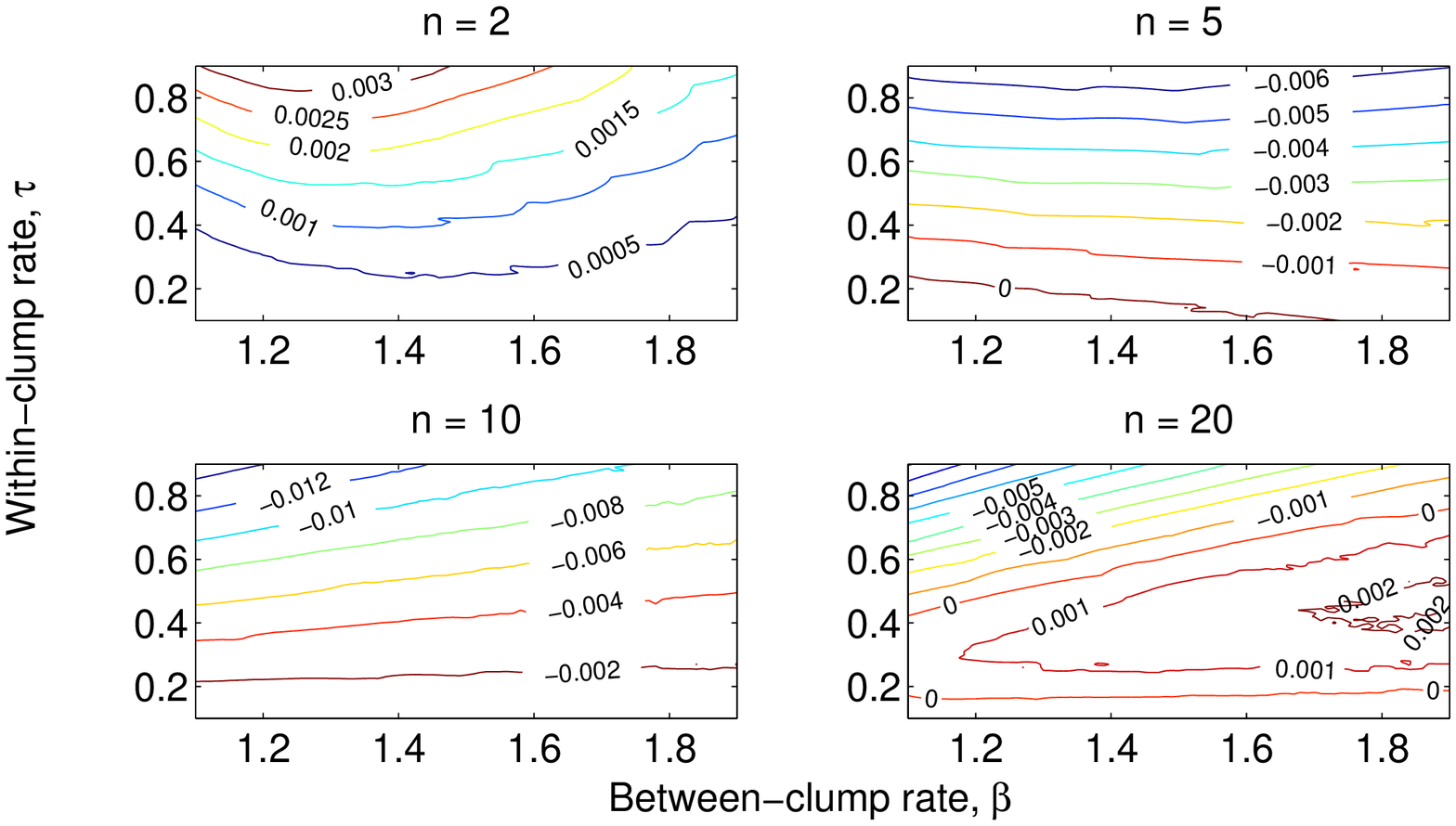} }}\\
\scalebox{0.75}{\resizebox{\textwidth}{!}{ \includegraphics{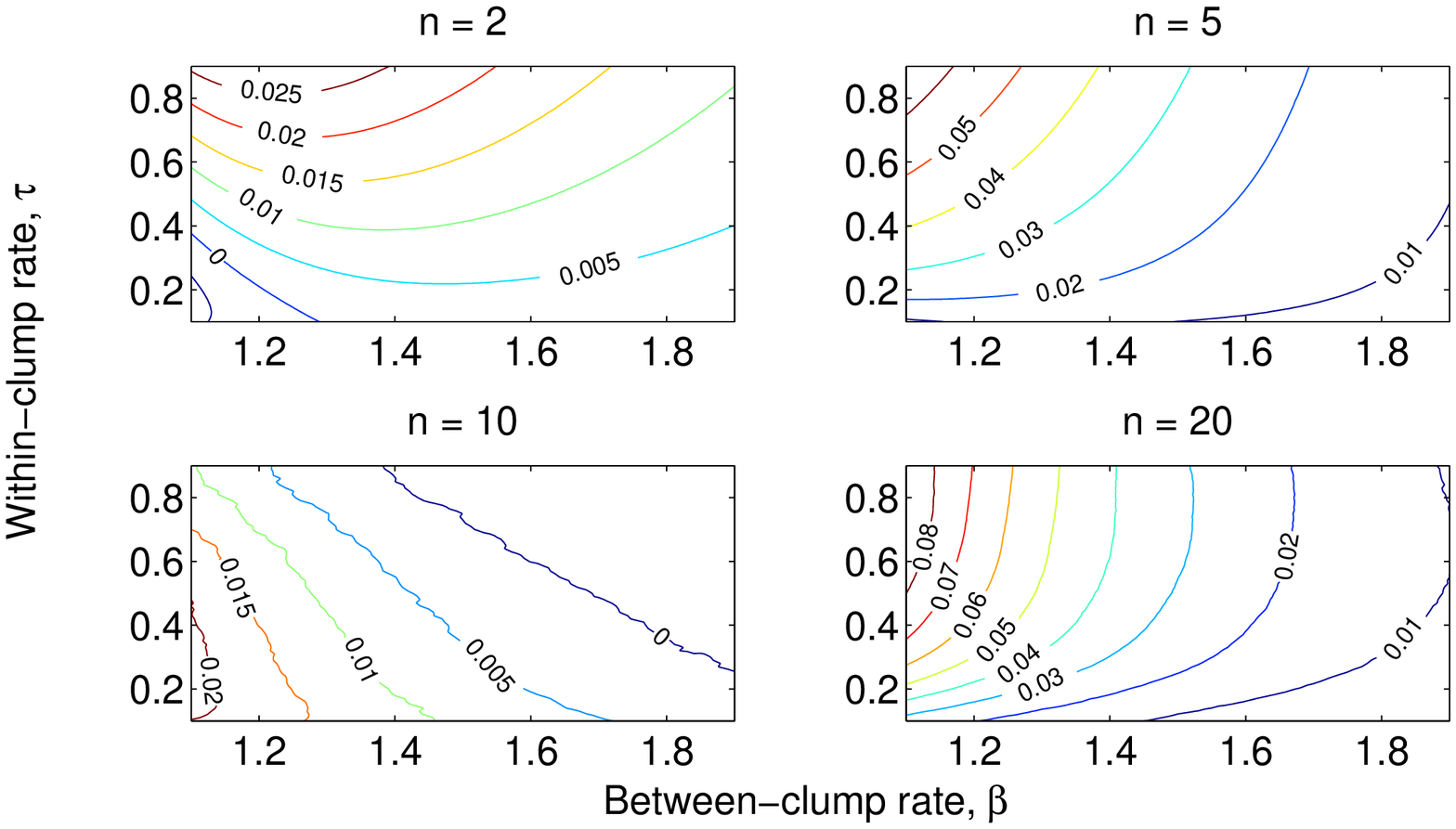} }}
\caption{Performance of Order-2 approaches for different values of $n$, $\beta$
and $\tau$, with $\gamma = 1$, (top) absolute error in peak (bottom) absolute
error in final proportion of the population susceptible. \label{scan}}
\end{figure}

These observations suggest a hybrid strategy for the SIR model since the
behaviour for small levels of infection is a linear dynamical system whose
dominant eigenvalue can be calculated efficiently \citep{Ross:2010}.
Figure~\ref{hyb} shows the results of switching from the early exponential
growth rate predicted by the linear system to the Order 2 system at various
switchover values of prevalence $J$, demonstrating that we can accurately
reduce the problem from 66 to 5 ODEs. Such a strategy could be generalised to
other population dynamics where a linear approximation can be used around the
fixed points, joined to algebraic methods used when non-linear effects become
important.

Finally, Figure~\ref{scan} shows the behaviour of the Order 2 system over a
range of values for $n$, $\tau$ and $\beta$ (since $\gamma$ can be eliminated
by a non-dimensionalisation it is not varied). These show that errors remain
relatively small, and that the approximation tends to work better for smaller
values of $\tau$. This is to be expected, since as $\tau \rightarrow 0$ both
the approximation and the full model tend to the standard SIR model.

\section{Macroparasite dynamics}

\label{macro}

Modelling infectious diseases caused by macroparasites such as worms has long
been a technical challenge in infectious disease
epidemiology~\citep{Anderson:1991,Anderson:1993}, with more recent impetus due
to the political prioritisation of mitigating neglected tropical
diseases~\citep{Anderson:2013,Addiss:2013}. Here each person is, from the point
of view of worm ecology, a local habitat analogous to the clumps considered
above.

Suppose we consider a model of macroparasite dynamics where $p_w(t)$ is the
proportion of people in the population with a worm burden of $w$. Individuals
clear worms at a rate of $d$ per worm, and experience a non-linear force of
infection from other infected individuals leading to dynamics
\be
\ddt{}{p}_{w} = d (-wp_w + (w+1)p_{w+1}) +
 \left( {\textstyle \sum_{w'} \lambda_{w'}p_{w'}} \right) (-p_w + p_{w-1}) \text{ .}
\label{wpeq}
\ee
These are a special case of the equations analysed by~\citet{Isham:1995} using
probability generating function and moment closure methods.  As before we will
write these using vector notation
\be
\ddt{\myvec{p}} = (d \mymat{D} + \Lambda(t) \mymat{E}) \myvec{p} \text{ ,}
\quad \text{where} \quad
\myvec{p}(t) := (p_{w}(t)) \text{ ,}
\quad
\Lambda(t) := \mygvec{\lambda} \cdot \myvec{p}(t) \text{ ,}
\quad
\mygvec{\lambda} := (\lambda_w) \text{ .} \label{wecdef}
\ee
It is then possible to see that the two matrices in this expression are
linearly independent and obey $[\mymat{D}, \mymat{E}] = - \mymat{E}$. They
therefore form a representation of a Lie algebra, meaning we can apply the general
methodology of \S{}\ref{sec:lie} above and expand
\be
\mymat{M}(t) = d \mymat{D} + \Lambda(t) \mymat{E} 
\text{ ,} \quad
\mymat{Z}(t) = {\delta(t) \mymat{D} + \varepsilon(t) \mymat{E}} 
\text{ ,} \quad
\mymat{G}(t,u) = \Delta(t,u) \mymat{D} + \mathcal{E}(t,u) \mymat{E} \text{ .}
\ee
Substituting these expansions into~\eqref{liedes} gives
\be
\frac{\partial \Delta}{\partial u}\mymat{D} +
\frac{\partial \mathcal{E}}{\partial u}\mymat{E}
= \left(\varepsilon(t) \Delta(t,u) - \delta(t) \mathcal{E}(t,u)\right)
\mymat{E} \text{ ,}
\quad
\Delta(t,0) \mymat{D} + \mathcal{E}(t,0) \mymat{E}
= \dot{\delta}(t) \mymat{D} + \dot{\varepsilon}(t) \mymat{E}
\text{ .}
\ee
From these equations, together with~\eqref{zint}, we obtain a solution
\be
\myvec{p}(t) = {\rm e}^{dt \mymat{D} + \varepsilon(t)\mymat{E}} 
\myvec{p}(0) \text{ ,} \quad
\ddt{\varepsilon} = \frac{\Lambda(t) - \varepsilon(t) \eta(t)}{\zeta(t)} \text{ ,} \quad
\zeta(t) := \frac{1-{\rm e}^{-dt}}{dt} \text{ ,} \quad \eta(t) :=
\frac{1-\zeta(t)}{t} \text{ .} \label{liesol}
\ee
Note that in contrast to the clump-based SIR model, this solution is exact.
There are two main benefits of the algebraic approach. First, while traditional
moment closures are based on the assumption that $p_w$ is the probability mass
function for some standard distribution -- typically the negative binomial --
we know here that the low-parameter form $\myvec{p} = {\rm e}^{\theta_1
\mymat{D} + \theta_2 \mymat{E}} \mygvec{\vartheta}$ holds exactly, giving a
similar level of model complexity to the negative binomial while being derived
directly from biological mechanisms. Secondly, since $w\in\{0,1,\ldots{}\}$ we
must choose a sufficiently large maximum $w$ and keep that fixed to work
with~\eqref{wpeq} directly. In contrast, for~\eqref{liesol} we have one ODE and
can select whatever maximum value of $w$ is required to secure accurate
calculation of the matrix exponential each time step.

\section{Discussion}

The approximate algebraic method proposed here is not dependent on special
features of SIR dynamics, and could be applied to other systems of population
dynamics -- such as more realistic epidemiological models (e.g.\ SEIR and
SIRS), voter models, or predator-prey systems -- provided self-consistent
equations like~\eqref{sc} are appropriate. The topology of the network of
interactions between individuals also need not be the simplest clump model, and
other local-global dynamics -- such as the dynamics of \citet{House:2010},
\citet{Volz:2011}, and \citet{Ma:2013}, and potentially dynamics on still more
general network structures such as those considered by \citet{Karrer:2010} --
can be considered in this framework.

In general, however, the problem of convergence of a matrix series is likely to
be the major limitation of algebraic methods -- formal convergence is
typically too conservative in terms of what will constitute a good
approximation.  This would motivate derivation of an exact expansion of the
kind provided for macroparasite dynamics as in \S{}\ref{macro} above and
elsewhere for other relatively simple kinds of population dynamics
\citep{House:2012,Shang:2012}. At present, the selection of an appropriate
linearly independent matrix basis set for more realistic dynamics has not been
possible, however hopefully more fundamental approaches to the
algebraic treatment of time-inhomogeneous Markov chains such as work by
\citet{Sumner:2013} will help to provide improved, exact, closures.

\section*{Acknowledgements}

Work funded by the UK Engineering and Physical Sciences Research Council
(EPSRC). The author would like to thank Joshua Ross, the editors, and three
anonymous reviewers for helpful comments on this manuscript.

\end{document}